
\input harvmac.tex      
\font\blackboard=msbm10 \font\blackboards=msbm7
\font\blackboardss=msbm5
\newfam\black
\textfont\black=\blackboard
\scriptfont\black=\blackboards
\scriptscriptfont\black=\blackboardss
\def\blackb#1{{\fam\black\relax#1}}
\def\BR{{\blackb R}}
\def\BZ{{\blackb Z}}
\def\BQ{{\blackb Q}}

\def\tq{\tilde{q}}
\def\tn{\tilde{n}}
\def\tPhi{\tilde{\Phi}}
\def\tphi{\tilde{\phi}}

\def\bpsi{\bar{\psi}}
\def\bz{\bar{z}}

\def\vev{\omega}
\def\tvev{\tilde{\omega}}
\def\Sh{\hat{S}}

\def\Pol{J. Polchinski, Nucl. Phys. {\bf B242} (1984) 345.}
\def\Bos{S. Coleman, Phys. Rev. {\bf D 11}, 2088 (1975);
    S. Mandelstam, Phys. Rev. {\bf D11}, 3026 (1975).}
\def\NT{I. Niven, H. S. Zuckerman, and H. L. Montgomery,
  {\it An Introduction to the Theory of Numbers}, 5th ed.
  (John Wiley \& Sons, Inc., 1991).}
\def\GSW{M. B. Green, J. H. Schwarz, and E. Witten,
  {\it Superstring Theory} (Cambridge University Press, 1988).}

\def\GR{I. S. Gradshteyn and I. M. Ryzhik,
  {\it Table of Integrals, Series, and Products}, Alan
  Jeffrey ed. (Academic Press, Inc., 1980).}


\Title{PUPT - 1455, March 1994}
{\vbox{\centerline{Fermions Coupled to a Conformal Boundary:}
\centerline{A Generalization of the Monopole-Fermion System} }}

\bigskip
\bigskip

\centerline{Ali Yegulalp\footnote{*}{
  yegulalp@puhep1.princeton.edu}}
\medskip\centerline{Joseph Henry Laboratories}
\centerline{Princeton University}
\centerline{Princeton, NJ 08544}
\bigskip\bigskip

\vskip .25in

We study
a class of models in which $N$ flavors of massless fermions on the
half line are coupled by an arbitrary
orthogonal matrix to $N$ rotors living on
the boundary.  Integrating out the rotors, we find the
exact partition function and Green's functions.
We demonstrate that the coupling matrix must satisfy
a certain rationality constraint, so there is an infinite,
discrete set of possible coupling matrices.
For one particular choice of the coupling matrix,
this model reproduces the low-energy dynamics of fermions
scattering from a magnetic monopole.
A quick survey of the Green's functions shows that
the S-matrix is nonunitary.  This nonunitarity is present
in previous results for the monopole-fermion system, although
it appears not to have been noted.  We indicate how
unitarity may be restored by expanding the Fock space
to include new states that are unavoidably introduced by
the boundary interaction.

\Date{}

\eject

\newsec{Introduction}
We consider the action
\eqna\fermact
$$
\eqalignno{
A & =\int dt \,\Biggl[\, {1\over 2} \sum_{i=1}^{N}{
  {\dot{\alpha}_{i} (t)}^2 \over I_{i}} +
  \sum_{i,j=1}^{N} {\alpha_{i} (t) M_{ij}
   \big(:\psi^{\dagger}_{j}(0,t) \psi_{j} (0,t) :
     - : \bar{\psi}^{\dagger}_{j}(0,t) \bar{\psi}_{j} (0,t) : \big)}
  \cr &
  + i \sum_{i=1}^{N} \int_{0}^{L} dx \, \Bigl(
  \psi^{\dagger}_{i}(x,t) (\partial_{t} - \partial_{x})
  \psi_{i}(x,t) +
  \bar{\psi}^{\dagger}_{i}(x,t) ( \partial_{t} + \partial_{x} )
  \bar{\psi}_{i}(x,t)
  \Bigr)
  \Biggr], &\fermact {} \cr}
$$
where $\psi_i$ and $\bar{\psi}_i$ are chiral components of
Dirac fermion
fields, $M_{ij}$ is an arbitrary real, orthogonal matrix,
 $\alpha_i$ are the
rotor degrees of freedom, and the $I_i$ are constants.
In writing the action, we have split
the Dirac fermions into their two single-component chiral
constituents, $\psi_i$ and $\bar{\psi}_i$, with the index $i$
labeling flavors.

The action describes N flavors of Dirac fermions
living on the interval $0 \le x \le L$.  There are no interactions
except at $x=0$, where the fermion currents are linearly coupled
to N rotor coordinates.  The interaction at $x=0$ dynamically
couples the left movers $\psi_i$ to the right movers
$\bar{\psi}_i$ through the rotor coordinates.  At $x=L$, we
impose the simple reflecting boundary condition
$\psi_i(L,t) = e^{2\pi i \lambda_i}\bar{\psi}_i(L,t)$
so that the system will
be closed and self-contained.

In general, the coupling at $x=0$ allows the fermions and the
rotors to exchange energy, but we will be interested purely
in the cases where energy is not exchanged:  namely, when
$I_i \to \infty$ or $I_i \to 0$.
If we take all the $I_i \to \infty$, the rotors decouple and
a simple reflecting boundary condition is imposed on the
fermion currents at $x=0$.  If we take some $I_i \to \infty$
and the others to zero, then things are much more interesting;
we will find that we
can integrate out the rotators, leaving behind
a non-trivial, conformally invariant boundary
condition on the fermions.  The particular boundary condition
obtained depends on which $I_i \to 0$ and the value of the
matrix $M_{ij}$.

An interesting special case arises when we chose $M_{ij}$ and
$I_i$ so that the action becomes essentially
equivalent\footnote*{Polchinksi uses right moving Weyl fermions
on the full line, while we consider right and left moving
Dirac fermions on
the half line.  One can convert the half-line Dirac theory
into a full-line Weyl theory by defining
$\bar{\psi}_i(-x,t) =  \psi_i(x,t)$ for $x>0$, so that
left movers at $x>0$ get reflected into right movers
at $x<0$. }
to one first considered
by Polchinski \ref\rPol{\Pol}.  Polchinski's action captures the essential
physics of charged fermions scattering from a magnetic monopole
in four dimensions.

\newsec{The Partition Function}
We use standard bosonization techniques \ref\rBos{\Bos} to rewrite the
action in
terms of the boson fields $\Phi_i$:
\eqna\boseact
$$\eqalignno{
A  =\int dt \,\Biggl[\, {1\over 2} \sum_{i=1}^{N}{
  {\dot{\alpha}_{i} (t)}^2 \over I_{i}} +
  \sum_{i,j=1}^{N} {\alpha_{i} (t) M_{ij}
  \dot{\Phi}_j(0,t)} \, +
  \cr
  {1\over {8 \pi}} \sum_{i=1}^{N}{
  \int_{0}^{L} dx \big(\dot{\Phi}_i(x,t)^2 -
    {\Phi}'_i(x,t)^2 \big)}
\Biggr],
  & &\boseact {} \cr}
$$
where $\dot{\Phi} = \partial_t\Phi$ and $\Phi' =
\partial_x\Phi$.
The bosons are compact, so $\Phi_i \equiv \Phi_i + 2\pi$.
We fix $\Phi_i(L,t) = \theta_i$,
where $\theta_i = \pi(1-2\lambda_i)$, and allow $\Phi_i(0,t)$ to vary
with the action.  The classical equations of motion
obtained by varying the action are
\eqna\classmo
$$\eqalignno{
\ddot{\alpha}_i(t) = I_i \sum_{j=1}^{N}{
       M_{ij}\dot{\Phi}_j(0,t)}  & \cr
\Phi'_j(0,t) - 4 \pi \sum_{i=1}^{N}{
       \dot{\alpha}_{i}(t) M_{ij}} = 0  & &\cr
\Phi''_i(x,t) - \ddot{\Phi}_{i}(x,t) = 0 & &\classmo{}  \cr}
$$
Eliminating the rotor coordinates $\alpha_i$ and
changing over to the `rotated' fields
$\tPhi_i(x,t) = \sum_{j=1}^{N}{M_{ij} \Phi_{j}(x,t)}$,
we get
\eqna\tildmo
$$\eqalignno{
\tPhi''_i(x,t) - \ddot{\tPhi}_i(x,t) = 0 & &\cr
4 \pi I_i \tPhi_i(0,t) + \tPhi'_i(0,t) - C_i = 0, & &\tildmo{} \cr}
$$
where the $C_i$ are constants of integration.

At this point, let us consider the limit $I_i \to \infty$
for all $i$, which should reproduce the free fermion
theory.  The bosons are
subject to the boundary conditions
$\Phi_i(0,t) = 0,\Phi_i(L,t) = \theta_i$ and the compactness
condition $\Phi_i(x,t) \equiv \Phi_i(x,t) + 2\pi$, which
result in the following standard free partition function:
\eqn\freebose{
Z = \prod_{n=1}^{\infty}{\big(1-q^n\big)^{-N}
    \prod_{i=1}^{N}{ \Biggl[
       \sum_{n=-\infty}^{\infty}
         {q^{{1 \over 2}(n+{\theta_i \over{2\pi}})^2}}}
       \Biggr]
     }
,}
where $q=e^{-\pi\beta/L}$ and $\beta=$ inverse temperature.
Note the presence of the factor
$\sum_{n=-\infty}^{\infty}
{q^{{1 \over 2}(n+{\theta_i \over{2\pi}})^2}}$,
which is simply a sum over winding number.  Without
the compactness condition on $\Phi_i$, this factor would
not be present.

For the fermion system, we have the boundary conditions
$\psi_i(0,t)=\bar{\psi}_i(0,t)$ and
$\psi_i(L,t)=e^{2\pi i\lambda_i}\bar{\psi}_i(L,t)$.
This yields the partition
function

\eqn\freeferm{
Z = \prod_{i=1}^{N}{
      \prod_{n = -\infty}^{\infty}
      {\Bigl(1+q^{|n-\lambda_i|}\Bigr)^2}
    }
.}

An application of Jacobi's triple product formula shows
that the bosonic and fermionic partition functions are indeed
equal up to a shift of zero-point energy,
verifying our choice of boundary conditions for the
bosonized action.

Now let us consider the general case $I_i \to \infty$
for $1 \le i \le a$ and  $I_i \to 0$ for
$a < i \le N$.  We find that
\eqna\eqmo
$$\eqalignno{
\tPhi''_i(x,t) - \ddot{\tPhi}_i(x,t) = 0,\qquad & 1 \le i \le N
  &\eqmo{a} \cr
\tPhi_i(0,t) = 0,\qquad & 1 \le i \le a &\eqmo{b}\cr
\tPhi_i'(0,t) = 0,\qquad & a < i \le N &\eqmo{c}\cr
\tPhi_i(L,t) = \sum_{j=1}^{N}{M_{ij}\theta_j},\qquad & 1\le i \le N
   &\eqmo{d}\cr
\tPhi_i(x,t) \equiv \tPhi_i(x,t) + 2\pi\sum_{j=1}^{N}{M_{ij}n_j},
    \qquad & n_j \in \BZ  &\eqmo{e}\cr
}
$$
In principle, there should be a constant of integration
in equation \eqmo{c}, but we have set it to zero because
the rotor coordinate will carry an infinite amount of energy
otherwise.

Since equations \eqmo{a} - \eqmo{e} describe a set of $N$
uncoupled free bosons, it seems that we should be able to compute the
partition function as a product of $N$ independent partition
functions. The fact that $M$ is an orthogonal matrix
means that the canonical commutation relations are
\eqn\comrel{
[\tPhi_i(x,t),\dot{\tPhi}_j(y,t)] =
[\Phi_i(x,t),\dot{\Phi}_j(y,t)] =
4\pi i \delta_{ij}\delta(x-y),
}
so there is no problem with quantizing the system in terms
of the $\tPhi_i$.
The only catch is that we must take care in
treating the winding modes.  When the boundary condition
at $x=0$ is $\tPhi'_i(0,t) = 0$, it turns out that no
winding mode exists, even though $\tPhi_i(x,t)$ is compact.
This lead us to the partition function
\eqn\zgeneral{
Z = \prod_{n=1}^{\infty}{
  \Bigl(1-q^n\Bigr)^{-a}
  \Bigl(1-q^{(n-{1\over 2})}\Bigr)^{-(N-a)}
  }
     \sum_{(\tilde{n}_1,\ldots \tilde{n}_a ) \in {\cal Z}_M}
     {q^{{1\over 2} \sum_{i=1}^{a} \big(
       \tilde{n}_i + \tilde{\theta}_i
        \big)^{2}
          }
      },
}
where
\eqna\defs
$$\eqalignno{
\tilde{\theta}_i &= \sum_{j=1}^N{M_{ij}{{\theta_j}\over{2\pi}} } \cr
{\cal Z}_M &= \biggl\{(\tilde{n}_1,\ldots \tilde{n}_a ) \biggm|
 \tilde{n}_i = \sum_{j=1}^N{M_{ij}n_j},\,\,\,
n_j \in\BZ \biggr\}. &\defs{} \cr
}$$

The partition function almost looks like that of a free theory,
except that the sum over winding modes is rather peculiar.
Barring the trivial cases $a=0$ and $a=N$, the winding mode
sum no longer factors into a product of independent
winding mode sums.  Instead, the winding modes of the different
boson flavors are coupled together by the matrix $M$.

Furthermore, it seems that the winding number sum in
the partition function does not
make sense unless we require $M$ to satisfy a certain rationality
property, the details of which will be explained in the
next section.
When $M$ satisfies the rationality property,
the set ${\cal Z}_M$ is a discrete lattice, so
points in ${\cal Z}_M$ are separated by finite gaps.  When
$M$ does not satisfy the rationality property, the set
${\cal Z}_M$ becomes dense on $\BR^{a}$.  Since we sum over
the points of ${\cal Z}_M$ in the winding number sum,
we find that the partition function diverges due to the
infinitesimally close spacing of energy levels.

To flesh out the substance of the preceding remarks,
we will work out a couple of simple examples:
the general $N=2$ case and the magnetic monopole
for arbitrary $N$.

\newsec{$N=2$}
Let us consider $N=2$, $a=1$, $\theta_i=0$,
and write $M$ in the form
\eqn\mtwo{
M =
 {1 \over {\sqrt{1+r^2}}}
    \left(\matrix{1 & r \cr
                  -r & 1 \cr}\right).
}

This gives us
\eqn\ztwo{
{\cal Z}_M = \biggl\{\tilde{n}\biggm|
   \tilde{n} = {{n_1+rn_2}\over{\sqrt{1+r^2}}};\,\,\,
   n_1,n_2 \in\BZ \biggr\}.
}

Using some elementary number theory \ref\rNT{\NT}, we see that
${\cal Z}_M$ becomes dense on $\BR$ unless $r$ is rational.
For $N=2$, the constraint $r \in \BQ$ is the rationality
condition for $M$.
Writing $r=p/l$ for $p,l\in\BZ$ with ${\rm gcf}(p,l)=1$,
we find that ${\cal Z}_M= {1\over{\sqrt{p^2+l^2}}}\BZ$, so the
partition function is
\eqn\ztwo{
Z = \prod_{n=1}^{\infty}{
  \Bigl(1-q^n\Bigr)^{-1}
  \Bigl(1-q^{(n-{1\over 2})}\Bigr)^{-1}
  }
  \sum_{m=-\infty}^{\infty}{
    q^{{m^2}\over{2(p^2+l^2)}}
  }.
}

Using standard techniques \ref\rGSW{\GSW},
 the partition function may be
reexpressed in terms of $\tq=e^{{4\pi^2}\over\ln(q)}$:
\eqn\ztwomod{
Z = \sqrt{{p^2+l^2}\over 2} \tq^{-{{1}\over{12}}}
    q^{1\over{48}}
  \prod_{n=1}^{\infty}{
  \Bigl(1-\tq^n\Bigr)^{-1}
  \Bigl(1-\tq^{(2n-1)}\Bigr)
  }
  \sum_{n=-\infty}^{\infty}{
     \tq^{{1\over{2}}(p^2+l^2)n^2}
  }
}

Now we can easily take the limit $L \to \infty$:
\eqn\lnztwo{
\ln(Z) \to {1\over 2}\ln({{p^2+l^2}\over 2}) +
           {{\pi L}\over{3\beta}} + O(e^{-L/2\beta}).
}
In addition to the standard piece in $\ln(Z)$ which scales
linearly in the size of the system, we see that there is
an $L$-independent term.  If we associate the size-independent
term with the boundary interaction at $x=0$, we find
that the boundary contributes a temperature-independent entropy
$S_{b}={1\over 2}\ln({{p^2+l^2}\over 2})$.  Equivalently,
this means that there are $g$ states associated with
the boundary, where $g=e^{S_b}=\sqrt{{p^2+l^2}\over 2}$.
The fact that $g$ need not be an integer may seem peculiar,
but it is an unavoidable consequence of the way the
winding modes for different bosons become linked
at the boundary.  The crucial point is that $g>1$, giving
us a hint that a correct treatment of scattering
may need to take into account some
hidden degree of freedom on the boundary.

\newsec{The Magnetic Monopole}
Now we take $N \ge 2$, choose $a=N-1$ and
$\theta_i = 0$, and set
\eqn\mmon{
M = \left(\matrix{
  {1\over{\sqrt{2}}} & -{1\over{\sqrt{2}}} & 0 & 0 & \ldots
     & 0 & 0\cr
  {1\over{\sqrt{6}}} & {1\over{\sqrt{6}}} & -{2\over{\sqrt{6}}}
     & 0 & \ldots & 0 & 0\cr
  \vdots & \vdots & \vdots & \vdots & \ddots & \vdots & \vdots\cr
  {1\over{\sqrt{N(N-1)}}} & {1\over{\sqrt{N(N-1)}}} &
     {1\over{\sqrt{N(N-1)}}} & {1\over{\sqrt{N(N-1)}}} &
     \ldots & {1\over{\sqrt{N(N-1)}}} &
     -{{(N-1)}\over{\sqrt{N(N-1)}}} \cr
  {1\over{\sqrt{N}}} & {1\over{\sqrt{N}}} &{1\over{\sqrt{N}}} &
    {1\over{\sqrt{N}}} & \ldots & {1\over{\sqrt{N}}} &
    {1\over{\sqrt{N}}} \cr
  }\right).
}
The last row of $M$ causes the sum of the fermion currents
to be  coupled to a rotor $\alpha_N$ which has $I \to 0$.
The first $N-1$ rows of $M$ couple to rotors with
$I \to \infty$, ensuring that differences
of fermion currents obey reflecting boundary conditions
at $x=0$. In Polchinski's version \rPol, there is just a single
rotor, which correponds to our $\alpha_N$.  The essence
of the model is that we are changing only the boundary condition
on the current which carries the total $U(1)$ charge.

The only bit of work we need to perform is to find
a convenient way to make the sum over ${\cal Z}_M$
explicit.  If we think of the numbers $\tn_i$ as functions
$\tn_i(n_1,n_2,\ldots,n_N) = \sum_{j=1}^{N}{M_{ij}n_j}$, then we see
that $\tn_i(n_1,n_2,\ldots,n_{N-1},n_N+1) =
\tn_i(n_1-1,n_2-1,\ldots,n_{N-1}-1,n_N)$ for
$1 \le i \le N-1$, so we can fix
$n_N=0$ and just sum over $n_1,n_2,\ldots, n_{N-1}$.
Note that we don't care about the value of $\tn_N$
since it does not appear in the winding mode sum.
Furthermore, summing over $n_1,n_2,\ldots,n_{N-1}$
gives us each possible value of $\tn_i$ exactly
once;  this is true because the matrix obtained by
deleting the last row and column of $M$ is a
non-singular matrix.  Finally, we note that
$\sum_{i=1}^{N-1}{\tn_i^2} =
\sum_{i=1}^{N}{n_i^2} - {1\over{N}}\big(\sum_{i=1}^{N}{n_i}\big)^2$
  Our desired partition function is
\eqn\zmon{
Z = \prod_{n=1}^{\infty}{
  \Bigl(1-q^n\Bigr)^{-(N-1)}
  \Bigl(1-q^{(n-{1\over 2})}\Bigr)^{-1}
  }
    \sum_{{n_i=-\infty}\atop{1\le i \le N-1}}^{\infty}{
       q^{{1\over{2}}\big(
          \sum_{i=1}^{N-1}{n_i^2} -
           {1\over{N}}\big(\sum_{i=1}^{N-1}{n_i}\big)^2
           \big)
          }
     }.
}

As in the $N=2$ case, we can express $Z$ in terms of
$\tq = e^{{4\pi^2}\over{\ln(q)}}$:
\eqn\zmonmod{
Z=
 {  {
    \tq^{-{N\over{24}}}
    q^{{N\over{24}}-{1\over{16}}}
    }
   \over
    {\sqrt{
       2^N
       \det(A)
      }
    }
 }
  \prod_{n=1}^{\infty}{
      \Bigl(1-\tq^n\Bigr)^{-(N-1)}
      \Bigl(1-\tq^{2n-1}\Bigr )
   }
    \sum_{{n_i=-\infty}\atop{1\le i \le N-1}}^{\infty}{
       \tq^{ {1\over{4}}
       \sum_{i,j=1}^{N-1}{(A^{-1})_{ij}n_in_j}
       }
    },
}
where $A_{ij} = -{1\over{2N}} + {1\over{2}}\delta_{ij}$.
$\det(A)$ may be evaluated using a standard formula
for circulants \ref\rGR{\GR}, yielding
$\det(A) = 2^{1-N}{1\over{N}}$.  Taking $L\to \infty$,
we find the boundary degeneracy to be
$g = \sqrt{N\over{2}}$.  Note that $g=1$ when $N=2$, so
we should expect that there are no degrees of freedom on
the boundary when $N=2$.

\newsec{Green's Functions}
Now we set $L=\infty$ and proceed to
compute
an arbitrary fermionic Green's
function $\Gamma$:
\eqn\fgreen{
\Gamma = <0|\bpsi_{i_1}(\bz_1) \ldots \bpsi^{\dagger}_{i_p}(\bz_p)
  \psi_{j_1}(w_1) \ldots \psi^{\dagger}_{j_q}(w_q)|0>_B,
}
where we have switched to holomorphic coordinates
$z = \tau + ix$ and imaginary time $\tau= it$.
To distinguish between the presence of $\psi$ and
$\psi^{\dagger}$, we assign $F_k=1$ for $\bpsi_{i_k}$,
$F_k=-1$ for $\bpsi^{\dagger}_{i_k}$, $G_k=1$ for $\psi_{j_k}$,
and $G_k=-1$ for $\psi^{\dagger}_{j_k}$.  The notation
$<...>_B$ indicates an expectation value in the presence
of the boundary interaction at $x=0$, while
$<...>$ will be used to indicate an expectation value
for a free theory on the full line with no boundary interaction.

As in the finite volume case,
we bosonize the system according to the
standard correspondence $\psi_i(z) = e^{i\Phi^L_i(z)}$
and $\bpsi_i(\bz) = e^{-i\Phi^R_i(\bz)}$
\footnote*{To avoid cumbersome
anticommuting factors normally appearing in multi-flavor
bosonization, we simply assume that $\Gamma$ is
written with the fields in the canonical ordering
$i_1 \le i_2 \le \ldots i_p$,
$j_1 \le j_2\le \ldots j_q$.
}.
Defining $\tPhi_i(z,\bz) = \sum_{j=1}^{N}{M_{ij}\Phi_j(z,\bz)}$,
we find that
\eqna\tPhimo
$$\eqalignno{
\del_z\del_{\bz}\tPhi_i(z,\bz) &=0 \qquad \hbox{for } \Im(z) > 0 \cr
\tPhi_i(z,\bz) &= 0 \qquad \hbox{for } 1 \le i \le a,\,\, z=\bz \cr
(\del_z-\del_{\bz})\tPhi_i(z,\bz) &=0 \qquad \hbox{for } a < i \le N,
  \,\,z=\bz. &\tPhimo{}\cr
}
$$

We can decompose $\tPhi_i(z,\bz)$ into left and right moving
fields $\tPhi^L_i(z)$ and $\tPhi^R_i(\bz)$, but
we must remember that the decomposition is not unique
since $\tPhi_i(z,\bz) = \tPhi^L_i(z) + \tPhi^R_i(\bz) =
(\tPhi^L_i(z) + C) + (\tPhi^R_i(\bz) - C)$.
Accordingly, let us define constants $\tvev_i$ such that
$<\tPhi^L_i(z)> = -<\tPhi^R_i(\bz)> = \tvev_i$.
For convenience, define $\tphi^L_i(z) = \tPhi^L_i(z) -\tvev_i$
and $\tphi^R_i(\bz) = \tPhi^R_i(\bz) + \tvev_i$.
Using equation \tPhimo{}, we can solve for the left
movers in terms of the right movers:
\eqna\lr
$$\eqalignno{
\phi^L_i(z) &= \sum_{k=1}^{N}{S_{ik}\phi^R_k(z)} \cr
S_{ik} &= -\sum_{j=1}^{N}{\sigma_j M_{ji}M_{jk}} \cr
\sigma_i &=\cases{1,&$1\le i \le a$\cr
                  -1,&$a<i\le N$ \cr}, &\lr{} \cr
}
$$
where $\phi_i = \sum_{j=1}^{N}{M_{ji}\tphi_j}$.

The matrix $S$ is manifestly symmetric, and a short calculation
shows that it is also orthogonal.  Furthermore, any matrix
which is real, symmetric, and orthogonal may be written
in the form $S_{ik} = -\sum_{j=1}^{N}{\sigma_j M_{ji} M_{jk}}$
with $\sigma_j^2 =1$ and $M$ real and orthogonal.
{}From this, we see that the boundary interaction at $x=0$
merely `rotates' the left movers with respect to the
right movers, and we may choose the boundary interaction
so that any particular symmetric and orthogonal matrix
$S$ carries out the rotation.

In bosonized form, equation \fgreen{} becomes
\eqn\fbose{
\Gamma =
<0|e^{-iF_1\Phi^R_{i_1}(\bz_1)} \ldots
    e^{-iF_p\Phi^R_{i_p}(\bz_p)}
    e^{iG_1\Phi^L_{j_1}(w_1)} \ldots
    e^{iG_q\Phi^L_{j_q}(w_q)}
   |0>_B.
}
Using equation \lr{} to write all the $\Phi^L_i(z)$ in terms
of the $\Phi^R_i(z)$, we find that
$\Gamma$ is simply the expectation value of a product
of right moving chiral vertex operators.
Once $\Gamma$ is expressed in terms of just right moving
fields, we can forget about the presence of the boundary
interaction and compute as we would in a free theory.
Use standard techniques for vertex operators \rGSW,
we get
\eqna\general
$$\eqalignno{
\Gamma =&
  e^{i\sum_{r=1}^{N}{\vev_r(f_r+g_r)}}
  \prod_{r=1}^{N}{\delta_{f_r,\sum_{r'=1}^{N}{S_{rr'}g_{r'}}}}
  \prod_{{\alpha,\beta=1}\atop{\alpha < \beta}} ^ {p} {
     \big(\bz_{\alpha} - \bz_{\beta}\big)^{
           F_{\alpha}F_{\beta}\delta_{i_{\alpha}i_{\beta}}
                                          }
   }
  \prod_{{\lambda,\nu=1}\atop{\lambda < \nu}} ^ {q} {
     \big(w_{\lambda} - w_{\nu}\big)^{
           G_{\lambda}G_{\nu}\delta_{j_{\lambda}j_{\nu}}
                                          }
   }
\cr
&
  \prod_{\alpha=1}^{p}{
    \prod_{\lambda=1}^{q}{
      \big(\bz_{\alpha}-w_{\lambda}\big) ^{
         -F_{\alpha}G_{\lambda}S_{i_{\alpha}j_{\lambda}}
      }
    }
  },&\general{}\cr
}
$$
where $f_r = \sum_{\alpha=1}^{p}{F_{\alpha}\delta_{i_{\alpha}r}}$
and $g_r = \sum_{\lambda=1}^{q}{G_{\lambda}\delta_{j_{\lambda}r}}$.
Note that $f_r$ and $g_r$ are simply the total amount of each
flavor present in the ingoing and outgoing states, respectively.

For the magnetic monopole, $S_{ij} = -\delta_{ij} + {2\over{N}}$,
giving us
\eqna\gmon
$$\eqalignno{
\Gamma_{monopole} =
  e^{i\kappa\vev}
&
  \prod_{r=1}^{N}{\delta_{f_r+g_r,\kappa}}
  \prod_{{\alpha,\beta=1}\atop{\alpha < \beta}} ^ {p} {
     \big(\bz_{\alpha} - \bz_{\beta}\big)^{
           F_{\alpha}F_{\beta}\delta_{i_{\alpha}i_{\beta}}
                                          }
   }
  \prod_{{\lambda,\nu=1}\atop{\lambda < \nu}} ^ {q} {
     \big(w_{\lambda} - w_{\nu}\big)^{
           G_{\lambda}G_{\nu}\delta_{j_{\lambda}j_{\nu}}
                                          }
   }\cr
&
  \prod_{\alpha=1}^{p}{
    \prod_{\lambda=1}^{q}{
      \big(\bz_{\alpha}-w_{\lambda}\big) ^{
         F_{\alpha}G_{\lambda}(\delta_{i_{\alpha}j_{\lambda}}-{2\over{N}})
      }
    }
  },&\gmon{} \cr
}
$$
where $\kappa =
{2\over{N}}\sum_{r=1}^{N}{g_r}$ and $\vev =
\sum_{r=1}^{N}{\vev_r}$.
Our results for the monopole Green's functions agree
with the original calculation by Polchinski \rPol, in which
he evaluated the path integral by integrating out the
fermions.  The parameter $\vev$ in our result corresponds
to the vacuum instanton angle $\theta$ in Polchinski's result.
The path integral
calculation gives results which are integrated over
all values of $\theta$, from which Polchinski extracted
the $\theta$-vacuum Green's functions by cluster
decomposition arguments.
In our bosonized solution, we
automatically
obtain the Green's functions in a fixed $\theta$-vacuum
sector.

\newsec{A Unitarity Paradox}

Now we would like to point out that the Green's functions
of equation \gmon{} produce an apparently nonunitary
S-matrix for $N>2$\footnote*{
The unitarity problem is not particular to our treatment
of the monopole-fermion system;  the results obtained
in \rPol{} have exactly the same problem, although it
seems not to have been commented on until now.}.
Let us consider Green's
functions where we have some fixed combination of
left moving operators and any number of right moving
operators.  The idea here is that we fix an ingoing
state created by left moving operators, allow
scattering to take place, and then look at the overlap
with outgoing states made of arbitrary combinations
of right movers.  Since we have fixed the left movers,
the integer parameters $G_{\lambda}$ and $g_r$ are fixed, while
$F_{\alpha}$ and $f_r$ are free to vary.
Since $\kappa=
{2\over{N}}\sum_{r=1}^{N}{g_r}$,
we see that $\kappa$ is fixed, but not necessarily integral
for $N>2$.
Looking at equation \gmon{}, we see that the Green's function
will vanish unless $f_r+g_r=\kappa$ for all $r$, leading us
to the following conclusions:
for $N=2$, every choice of ingoing state has overlap with
some outgoing states, and there is no problem with unitarity.
In fact, the interaction at $N=2$ simply swaps the two flavors
upon reflection from the boundary.
For $N>2$, we find that any ingoing state with
a non-integer $\kappa$ (i.e., a state whose total charge
is not a multiple of $N/2$) has zero overlap with all possible
outgoing states, violating probability conservation and unitarity.

As a first step toward resolving the unitarity problem,
let us think about the S-matrix in terms of bosons.
The  advantage of the bosonized form is
that we can think of the scattering directly in terms
of operators:  by virtue of equation \lr{}, we can
think of the boundary S matrix as an operator
which maps the chiral, left moving `in' state operators into
the chiral, right moving `out' state operators.  Explicitly,
we have $\Sh\phi^L_i(z)\Sh^{-1} =
\sum_{j=1}^{N}{S_{ij}\phi^R_j(z)}$, where $\Sh$
is the Hilbert space S-matrix operator and we regard
$\phi^L_i(z)$ and $\phi^R_i(z)$ as operators
in a free theory without a boundary interaction.

Formally, it is easy to show that $\hat{S}$ acts on the
bosonic Fock space in a unitary way.
Let $|in>$ and $|in'>$ denote two arbitrary ingoing
states formed by acting on the vacuum $|0>$ with arbitrary
combinations of operators $\phi^L_i(z)$.  To demonstrate
unitarity, we need to show that
$<in'|\Sh^{\dagger}\Sh|in> = <in'|in>$.
If we let $\hat{S}$ and $\hat{S}^{\dagger}$ act on the
left moving fields in $|in>$ and $<in'|$, the left movers
turn into right movers multiplied by factors of $S_{ij}$.
Since we are now dealing with free field theory, the inner
product may be evaluated in terms of
two-point functions by Wick's theorem.  The orthogonality
of $S_{ij}$ implies that
${<0|\phi^L_i(z)\Sh^{\dagger}\Sh\phi^L_j(w)|0>} =
{\sum_{k,l=1}^{N}{S_{ik}S_{jl}<0|\phi^R_k(z)\phi^R_l(w)|0>}} =
{<0|\phi^L_i(z)\phi^L_j(w)|0>}$, which
finishes the proof that $\Sh$ is unitary.

Now let us consider what happens when $\Sh$ acts on fermionic
states.  Setting $N=4$ for simplicity, we find that
\eqna\sferm
$$\eqalignno{
\Sh\psi_1(z)\Sh^{-1} = \Sh e^{i\Phi_1^L(z)} \Sh^{-1} &=
e^{-{i\over 2} \Phi_1^R(z)}
e^{{i\over 2} \Phi_2^R(z)}
e^{{i\over 2} \Phi_3^R(z)}
e^{{i\over 2} \Phi_4^R(z)}
&\cr
\Sh\psi_2(z)\Sh^{-1} = \Sh e^{i\Phi_2^L(z)} \Sh^{-1} &=
e^{{i\over 2} \Phi_1^R(z)}
e^{-{i\over 2} \Phi_2^R(z)}
e^{{i\over 2} \Phi_3^R(z)}
e^{{i\over 2} \Phi_4^R(z)},
&\sferm{}\cr
}
$$
with similar expressions for the other flavors.
Every ingoing fermion operator scatters into a product
of vertex operators carrying half-integer charges of each
flavor.  Since the expectation value of
a product of vertex operators vanishes
unless there is a net charge of zero for each flavor,
we see that ingoing states with an odd number
of fermions have no overlap with any outgoing states
made of fermions.
This suggests a simple solution to our unitarity problem:
we should expand our Fock space and
allow
outgoing states to contain vertex operators carrying
half-integer charges.  The new operators are in fact
nothing but spin operators for our Dirac fermions.
The complete Fock space will be generated from fermion
operators and spin operators, and the boundary interaction
will couple the two kinds of operators together.
We can also think of the Fock space as being built from
fermion operators acting on two kinds of vacuum states:
one is the ordinary Raymond vacuum, and the other
is the Neveu-Schwarz vacuum created by a spin operator
acting on the Raymond vacuum.
When an even number of fermions scatter from the boundary,
we remain in the same vacuum.  When an odd number of fermions
scatter from the boundary, the outgoing states emerge
in the opposite vacuum from the ingoing states.  In some
sense, we can think of the boundary as having a two-fold
degeneracy corresponding to the two possible vacua.
We plan to present a more detailed exposition of these ideas
in a forthcoming work.

\newsec{Conclusions}
In this letter we have generalized the monopole-fermion
system to a class of conformal systems with conformal
boundary interaction parameterized by an orthogonal matrix.
Abelian bosonization allows completely explicit computation
of the partition function and
Green's functions, which agree with previous results obtained
for the monopole-fermion system.  We point out that
the scattering appears to be nonunitary when $N>2$, and suggest how
this may be corrected by recognizing the presence of
additional states of the boundary.
The partition function tells us that the boundary degeneracy is
$g = \sqrt{N\over{2}}$, so extra states for the boundary
are required precisely when $g>1$.

\newsec{Acknowledgements}

We would like to thank C. G Callan, A. W. W. Ludwig,
I. Klebanov, and J. Maldacena for useful discussions
and advice.  This work was supported in part by
DOE grant DE-FG02-90ER40542.

\listrefs

\end